\documentclass[conference]{IEEEtran}
\IEEEoverridecommandlockouts
\usepackage{cite}
\usepackage{balance}
\usepackage{comment}
\usepackage{amsmath,amssymb,amsfonts}
\usepackage{algorithmic}
\usepackage{subcaption}
\usepackage{graphicx}
\usepackage{caption}
\usepackage{booktabs}
\usepackage{float}
\usepackage{subcaption}
\usepackage{stfloats}
\usepackage{threeparttable}
\usepackage{booktabs}
\usepackage{orcidlink}
\usepackage{url}
\usepackage[utf8]{inputenc}
\usepackage[colorinlistoftodos]{todonotes}
\usepackage{multirow}
\usepackage{tabularx}
\usepackage{tikz}

\usepackage{textcomp}
\usepackage{xcolor}
\def\BibTeX{{\rm B\kern-.05em{\sc i\kern-.025em b}\kern-.08em
    T\kern-.1667em\lower.7ex\hbox{E}\kern-.125emX}}
\begin{document}

\title{Analyzing Nursing Assistant Attitudes Towards Empathic Geriatric Caregiving Using \\Quantitative Ethnography
}

\author{\IEEEauthorblockN{\textsuperscript{} Behdokht Kiafar} 
\IEEEauthorblockA{\textit{University of Delaware} \\
\textit
Newark, DE \\
\orcidlink{0009-0001-4415-1332}}
\and
\IEEEauthorblockN{\textsuperscript{} Salam Daher}
\IEEEauthorblockA{\textit{New Jersey Institute of Technology} \\
\textit
Newark, NJ \\
\orcidlink{0000-0002-2793-6293}}
\and

\IEEEauthorblockN{\textsuperscript{} Shayla Sharmin}
\IEEEauthorblockA{\textit{University of Delaware} \\
\textit
Newark, DE \\
\orcidlink{0000-0001-5137-1301}}
\and
\IEEEauthorblockN{\textsuperscript{} Asif Ahmmed}
\IEEEauthorblockA{\textit{New Jersey Institute of Technology} \\
\textit
Newark, NJ \\
\orcidlink{0000-0002-3108-138X}}
\and
\IEEEauthorblockN{\textsuperscript{} Ladda Thiamwong}
\IEEEauthorblockA{\textit{University of Central Florida} \\
\textit
Orlando, FL \\
\orcidlink{0000-0001-8506-5812}}
\and
\IEEEauthorblockN{\textsuperscript{} Roghayeh Leila Barmaki}
\IEEEauthorblockA{\textit{University of Delaware} \\
\textit
Newark, DE \\
\orcidlink{0000-0002-7570-5270}}
}


\maketitle

\begin{abstract}
An emergent challenge in geriatric care is improving the quality of care, which requires insight from stakeholders. Qualitative methods offer detailed insights, but they can be biased and have limited generalizability, while quantitative methods may miss nuances. Network-based approaches, such as quantitative ethnography (QE), can bridge this methodological gap. By leveraging the strengths of both methods, QE provides profound insights into need-finding interviews. In this paper, to better understand geriatric care attitudes, we interviewed ten nursing assistants, used QE to analyze the data, and compared their daily activities in real life with training experiences. A two-sample t-test with a large effect size (Cohen’s \textit{d}=$1.63$) indicated a significant difference between real-life and training activities. The findings suggested incorporating more empathetic training scenarios into the future design of our geriatric
care simulation. The results have implications for human-computer interaction and human factors. This is illustrated by presenting an example of using QE to analyze expert interviews with nursing assistants as caregivers to inform subsequent design processes.
\end{abstract}


\section{Introduction}

The number of people aged 85 and above is expected to double by 2036 and triple by 2049. However, there are already reports of staff shortages in assisted living and nursing homes \cite{harrington2020appropriate}. A significant barrier to care concerns perceptions and attitudes related to working with older adults. Caregivers play an enormous role in aging care, but inadequate training limits their effectiveness and affects their attitudes toward working with seniors. In addition to a need for appropriate training, nursing assistants must be able to communicate effectively with older patients \cite{campbell2020interventions}.

Nursing assistant training is a critical educational pathway for nursing students, guiding them from the early stages of learning to their future careers and preparing them for their professional roles, including those in geriatric care \cite{blay2020systematic}. Currently, the formal training for nursing assistants as caregivers incorporates role-playing. Despite their benefits, role-playing methods fall short of offering in-depth, realistic experiences to caregivers during their training. Simulation methods, on the other hand, may better prepare trainees for direct interactions in a realistic clinical setting. Recognizing the need to develop practical skills such as critical thinking, communication, flexibility, and empathy, experts suggest that we need to create and implement learning environments to better prepare caregivers \cite{wueste2023collaborate}. In response to this need, our paper introduces quantitative ethnography (QE) as a methodological approach to bridge the gap between theoretical knowledge and practical skills in geriatric caregiving. By analyzing need-finding interviews \cite{patnaik1999needfinding}, we reveal the meaningful connections between key elements of care, aiming to use this information to design more informed, clinically-relevant solution for caregiving simulation.

Our ultimate goal is to improve caregiver perceptions of working with older adults by developing an immersive simulation featuring a virtual geriatric patient. For this purpose, we need to identify the activities and challenges appropriate for nursing assistants to encounter in everyday interaction with geriatric patients to improve their attitudes. Since interviews and personal stories as powerful need-finding tools offer rich insight into understanding caregiver-older adult interactions and identifying workplace challenges, we interviewed ten caregivers in hour-long sessions and analyzed the data.

This paper aims to advance our understanding of caregiver-older adult interactions \cite{lunney2003critical, murray2019new}, by examining their attitudes, perceptions of older adults, and communication preferences using non-traditional, network-based analytical methods.

While quantitative methods efficiently identify data patterns for analyzing the interviews, they may overlook details. Qualitative methods provide deep and detailed insights that can be biased and less generalizable. Quantitative ethnography, on the other hand, is a methodological approach that combines the strengths of qualitative and quantitative methods to better understand and model complex human behaviors and their relationships with visual representations and illustrations. Despite being a powerful analytical tool, the number of studies leveraging QE in health informatics and human factors field, especially for need-finding data analysis is underrepresented. So, the key contribution of this paper is to introduce a working example of using QE to interpret interview data from nursing assistants for subsequent decision making and design guidelines by providing insights into the key characteristics of an ideal simulation tool for geriatric caregiving. 

We conducted semi-structured interviews with ten trained caregivers. After transcribing and coding the data, we used Epistemic Network Analysis (ENA) as a type of QE to examine the differences between their real-life caregiving experiences and what they learned in training.
We selected ENA to guide our analytical approach because of its two distinct advantages: 1) it visually represents concept connections through a graphical model and 2) offers robust statistical analysis to compare the code connections between models\cite{shaffer2016tutorial}. The analysis revealed a notable lack of empathy in formal training compared to the lived experiences of caregivers. The findings suggest a need for more empathetic training scenarios in our future geriatric care simulations. Our contributions are as follows:
\begin{itemize}
    \item Conducting need-finding semi-structured interviews with trained caregivers as subject matter experts.
    \item Providing a working example of the processed dataset with step-by-step guides to understand Epistemic Network Analysis process for our healthcare interview dataset.
    \item Reporting and discussing the results of Epistemic Network Analysis to provide insights about meaningful connections between important aspects of caregiving, especially centering on empathetic and communicative training, beneficial for healthcare decision-making, planning, and design of future systems.
\end{itemize}

The paper is structured as follows: We begin by explaining the role of caregivers and the importance of their training. Next, we explore the crucial aspects of caregiving. Then we discuss the theoretical foundations of ENA, along with the procedures used for data collecting and coding. After analyzing the data and presenting the results, we conclude in the final section.

\section{Background}

\subsection{Role of Caregivers}

Caregivers play a pivotal role in providing essential physical, emotional, and sometimes medical support, particularly for older adults, to address their various needs \cite{sabo2021self}. These supportive functions include assisting with daily activities, such as grooming, dressing, and meal preparation, all aimed at ensuring the overall well-being of older individuals \cite{reinhard2019home}. Caregivers also provide companionship, facilitate social interaction, and offer emotional comfort, which can significantly affect the mental and emotional health of older people \cite{murfield2020self}. 
Their role becomes essential in preserving the quality of life and independence of older adults, particularly those who may face challenges due to age-related conditions or disabilities \cite{wolff2020family}.
Effective communication between caregivers and individuals is another crucial aspect of delivering supportive care \cite{faronbi2019caring}. Establishing a dynamic and effective relationship fosters understanding, trust, and empathy, empowering assistants to meet patient needs and assist with daily care routines. \cite{singh2015training}.

Therefore, the role of caregivers is critical in ensuring the holistic well-being of older adults by providing a comprehensive range of support and addressing their various needs.

\subsection{Training for Caregivers}

Nursing assistant training includes a theoretical component (e.g., fundamental concepts for roles and responsibilities, ethics, communication, anatomy and physiology, safety, and providing care) and a laboratory component with in-person skills, after which they work directly with patients in the clinical setting \cite{redcross2022}. Caregivers have repeatedly expressed a lack of preparation and training in their caregiving roles, as this training is task-oriented and does not include attitudes-related training \cite{emich2019, blazer2016hearing}. High levels of emotional intelligence and various skills related to patient care are essential for this demanding role. Therefore, specialized training is required. Recent findings indicate that assisting family caregivers, including education and training, can potentially enhance the well-being of caregivers and individuals under their care \cite{blazer2016hearing, burgdorf2019factors}. Thus, a nursing program can equip them with the necessary skills, knowledge, and tools to perform their duties effectively and compassionately \cite{aksoydan2019training}.

These skills are essential for providing safe and effective care. During training, caregivers receive education about the specific conditions and needs of the people they care for \cite{gonzalez2021remotely}. Moreover, it includes instruction on providing emotional support and creating an encouraging environment for the care recipient. Meanwhile, caregivers learn techniques to manage stress, provide companionship, and engage in empathetic communication \cite{kamachi2020training}. Beyond traditional training methods, technology-based approaches have the potential to develop the clinical readiness of nursing students; however, several challenges need to be addressed to ensure their effective implementation \cite{lindeman2020technology}, such as simulating realistic caregiver-patient interactions and challenges.

\subsection{Key factors in Caregiving}
In the following, we discuss a few critical factors and elements highlighted by previous research \cite{wueste2023collaborate, bramhall2014effective, wu2021empathy} for effective caregiving and nursing responsibilities. The methodology will further use these elements to code the interview data for in-depth, quantitative ethnographic analysis. 

\subsubsection{Communication}

Communication is an essential part of the day-to-day elements of caregivers. Effective communication provides compassionate nursing care, which can improve the quality of care and emotional well-being of older adults \cite{bramhall2014effective}. Communication is considered a core competency as identified by the Accreditation Council for Graduate Medical Education and the American Board of Medical Specialties \cite{acgme2008}. The National Institute of Aging provides tips for communicating with older adults, such as speaking with them as fellow adults, making them comfortable, avoiding hurrying them, speaking plainly, addressing face to face, writing down or printing out key takeaways, and recognizing that people have different backgrounds and expectations  \cite{nianih}.
Some communication strategies are more effective than others; for example, simple sentences are more effective than slow speech \cite{ Small2003effectiveness}. 

\subsubsection{Empathy}
Empathy plays a crucial role in the education of nursing students, enabling them to provide effective care. It is central to the role of nursing assistants and is crucial for helping them to learn how to listen, communicate, and make decisions. An empathetic relationship between nursing assistants and older patients leads to positive health outcomes and job satisfaction \cite{ mirzaei2020effectiveness}. Understanding the concerns, experiences, and views of older patients may enhance the empathetic relationship and interaction that establish effective and healthy communication \cite{adler2000relationship}. A conversation analysis in nurse-patient interaction indicated that empathy can be achieved interactionally and sequentially in actual sequences of talk, and empathy is a collaboratively constructed action rather than the personal commitment of the caregiver \cite{ wu2021empathy}. There is a relationship between the cognitive and emotional components of empathy in nursing assistants \cite{ navarro2018engagement}. However, little work has been done to directly focus on the development of this skill of nursing assistants, especially using immersive simulations and virtual geriatric patients. 
\subsubsection{Flexibility}

Nursing assistants have a dynamic profession that involves caring for patients and individuals with diverse needs, medical conditions, and schedules \cite{mcvicar2003workplace}. Flexibility in caregiving includes a range of aspects, such as adapting to changing conditions, multitasking, providing emotional and psychological support, and managing unpredictable situations \cite{kirca2019relationship}. Patients come from diverse backgrounds and have unique care characteristics. As a result, caregivers must adapt and adjust their approaches to address individual patient needs, preferences, and cultural considerations \cite{sagbakken2020adapt}. In addition, nursing assistants often manage multiple patients and tasks simultaneously. Being flexible enables them to prioritize effectively, switch between tasks seamlessly, and manage their time efficiently to ensure all patients receive the necessary care \cite{kang2023toward}. Moreover, patients and their families may experience emotional distress, and caregivers need to be emotionally adaptable and responsive.
Overall, caregiving can be emotionally and physically demanding. Flexibility helps caregivers develop coping mechanisms and strategies to address the challenges associated with their profession and the unpredictable nature of healthcare environments \cite{palacio2020resilience}.

\subsubsection{Critical Thinking}

In geriatric care, caregivers need to employ critical thinking as it equips them to make wise decisions and provide a higher quality of care. Caregivers often face various obstacles and concerns related to the overall welfare of the individuals under their care. Critical thinking enables them to systematically examine and evaluate these issues, contemplate multiple potential resolutions, and ultimately select the optimal course of action aligned with the situation \cite{demiris2019problem}. Moreover, caregiving situations can change rapidly due to health fluctuations or unforeseen events. This skill allows caregivers to assess the new situation quickly, adjust their care strategies, and make informed decisions to provide the best possible support \cite{ploeg2020caregivers}.

In the following, we provide the theoretical framework of our analytic method, which employs the aforementioned elements of effective caregiving to conduct a thorough quantitative ethnographic analysis.

\subsection{Theoretical Foundations}
This section describes the theoretical foundation of our analytic tools: Quantitative Ethnography and Epistemic Network Analysis. These methodologies play a crucial role in identifying meaningful patterns from our need-finding interviews with caregivers.

\subsubsection{Quantitative ethnography (QE)} QE is a methodological framework that uses the detailed and contextual insights of qualitative methods in conjunction with quantitative techniques \cite{shaffer2017quantitative}. It allows us to not only discover the underlying patterns but also to visualize and quantify them. QE is one of the most illustrative ways to depict meaningful patterns by representing the structure of connections between the identified key concepts as a network of relationships.

\subsubsection{Epistemic Network Analysis (ENA)} ENA, as a component of the QE framework, serves as a specialized tool to measure and visualize connections between concepts \cite{shaffer2017quantitative}.
ENA provides a comprehensive framework for representing the interconnections among concepts and employs visual illustrations, mathematical frameworks, and statistical methods to examine quantifiable manifestations of qualitative data.

ENA operates on three essential principles.
First, it asserts the feasibility of identifying meaningful features within the data, which are represented as codes. These codes act as the identifiers or concepts for subsequent analysis. Secondly, ENA assumes the existence of local structures within the data. Thirdly, it focuses on examining the relationships between codes within these local structures \cite{bowman2021mathematical, shaffer2016tutorial, shaffer2017epistemic}. By quantifying the co-occurrence of codes within the data, ENA models the relationships between codes and generates a weighted network graph and associated visualizations for each unit of analysis.

Although initially developed to address learning analytics issues \cite{shaffer2009epistemic, fougt2018epistemic}, the ENA  analysis method can be expanded beyond learning \cite{d2021presentation, peters2019extending}. The core assumption of this method is that the structure of connections between the concepts is meaningful. In our interview data analysis, ENA was found to be a powerful tool because it effectively illustrated geriatric caregiver behaviors by modeling the relationships between their attitudes. These visual models, derived from real-life experiences, can inform caregiver training and align well with patient-centered care principles, ultimately elevating the healthcare training solutions.
\section{Methods and Materials}

\subsection{Data Collection}

\subsubsection{Participants} For this study, we conducted interviews with ten ($9$ females) participants aged between $19$ and $28$ ($M=22.71$ years) from Asian, White, and Hispanic ethnicities, all of whom had prior experience in nursing programs. Their professional caregiving experience ranged from five months to five years ($M=3\pm 2.2$ years). We recruited the participants by distributing flyers at the participating site, ensuring that they met the eligibility criteria, which required prior experience in caring for older adults. This approach to data collection ensures that our study is grounded in the real-world experiences of those directly involved in nursing education and practice, thereby strengthening the validity and applicability of our findings.

\subsubsection{Procedure} After participants expressed interest in participating in the interviews, they received the interview consent and prequestionnaire forms soliciting their demographics and background training/workforce experiences. After consent, we scheduled and held virtual interview sessions with participants via Zoom. The semi-structured interviews comprised 31 open-ended/follow-up questions, designed to encourage participants to provide more detailed and in-depth responses.
We asked participants about their regular job activities, interactions, dialogues with geriatric individuals/ patients, job incidents, and overall recommendations for a training simulation. 
The interview sessions lasted approximately an hour. All the sessions were recorded and subsequently transcribed using an automated transcription service called Trint \cite{trint2023}. Afterward, the research team reviewed and cleaned the automated transcription data to reduce potential errors. Once the data collection and preparation steps were completed, the transcription data was ready for in-depth analysis through coding and thematic analysis.

\begin{table} [t]
  \caption{Coding scheme for the analysis of geriatric caregiving expert interview transcriptions. Adapted from \cite{Small2003effectiveness, navarro2018engagement, danzl2016lot}.}
  \label{tab:coding-scheme}
  \begin{tabular} {lp{0.3\textwidth}}
    \toprule
   Code & Definition\\
    \midrule
    Communication &  Exchanging information, thoughts, feelings, and needs through verbal and non-verbal interactions.\\
    \\
    Empathy & Demonstrating a deep understanding and genuine concern for the emotional and psychological experiences of the patient.\\
    \\
    Flexibility & The ability to demonstrate adaptability, versatility, and openness in the care approach.\\
    \\
    Critical Thinking & The process of analyzing, evaluating, and interpreting information and situations related to the patient's health and treatment.\\
  \bottomrule
\end{tabular}
\end{table}

\subsection{Data formatting and Coding}

\begin{table*}[t]
\caption{Excerpt of coded data containing a few of interview questions and responses}
\label{tab:Excerpt}
\begin{threeparttable}
\begin{tabular}{lllllccccc}
\hline
\multicolumn{5}{c}{} 
& \multicolumn{4}{c}{}                                                                                                                                                     \\ 
\multicolumn{1}{l}{}     & \multicolumn{1}{c}{\textbf{Stanza}\tnote{1}}                                                                                                                                                                    & \multicolumn{2}{c}{\textbf{Unit}\tnote{2}}                                                                                                                                        & \multicolumn{1}{c}{\textbf{Raw Data}}                                                                                                                                                                                                                                                                                           & \multicolumn{4}{c}{\textbf{Codes}}                                                             \\ \hline
\multicolumn{1}{c}{\textbf{Line} } & \multicolumn{1}{c}{\textbf{Questions}}                                                                                                                                                                        & \multicolumn{1}{c}{\textbf{PID}\tnote{3}} & \multicolumn{1}{c}{\textbf{Category}\tnote{4}}                                                   & \multicolumn{1}{c}{\textbf{Response}}                                                                                                                                                                                                                                                                                           & \multicolumn{1}{c}{Communication} & \multicolumn{1}{c}{Empathy} & \multicolumn{1}{c}{Flexibility} & \begin{tabular}[c]{@{}c@{}}Critical \\ Thinking\end{tabular} \\ \hline
\multicolumn{1}{l}{1}    & \multicolumn{1}{l}{\begin{tabular}[c]{@{}l@{}}For residents 80 \\ up to 90, do they\\ have trouble with\\  hearing? What do \\ you have to say? \\ Can you give an \\ example of it?\end{tabular}}   & \multicolumn{1}{l}{RH}                                                        & \multicolumn{1}{l}{\begin{tabular}[c]{@{}l@{}}Real\\ Experience\end{tabular}}  & \begin{tabular}[c]{@{}l@{}}Yes, many of that age. I adjust by \\ speaking clearly, using visual aids, \\ and sometimes by writing to ensure \\ they understand. For example, other\\ day, I noticed a resident struggling\\ to hear, I gently repeated myself.\end{tabular} & \multicolumn{1}{c}{1}                                                         & \multicolumn{1}{c}{0}       & \multicolumn{1}{c}{1}           & 0                                                            \\ \hline
\multicolumn{1}{l}{2}    & \multicolumn{1}{l}{\begin{tabular}[c]{@{}l@{}}What makes a \\ patient/resident \\ easy to work \\with?\end{tabular}}                                                                                   & \multicolumn{1}{l}{RH}                                                        & \multicolumn{1}{l}{\begin{tabular}[c]{@{}l@{}}Real \\ Experience\end{tabular}} & \begin{tabular}[c]{@{}l@{}}A mutual understanding and respect. \\Being able to connect with them on\\ a personal level, understanding their \\ feelings, and making them feel heard \\and valued.\end{tabular}                                                                     & \multicolumn{1}{c}{1}                                                         & \multicolumn{1}{c}{1}       & \multicolumn{1}{c}{0}           & 0                                                            \\ \hline
\multicolumn{1}{l}{3}    & \multicolumn{1}{l}{\begin{tabular}[c]{@{}l@{}}In previous training, \\ how much of the \\ scenario would \\ focus on activities \\ versus \\ communication?\end{tabular}}                            & \multicolumn{1}{l}{CM}                                                        & \multicolumn{1}{l}{Training}                                                   & \begin{tabular}[c]{@{}l@{}}More focused on how to do the high \\standard of care and safety versus\\ how to build a relationship.\end{tabular}                                                                                                                                       & \multicolumn{1}{c}{0}                                                         & \multicolumn{1}{c}{0}       & \multicolumn{1}{c}{0}           & 1                                                            \\ \hline
\multicolumn{1}{l}{4}    & \multicolumn{1}{l}{\begin{tabular}[c]{@{}l@{}}What types of\\  activities did \\ you participate\\ during your \\ training? Could\\ you provide \\ an example of \\ one such activity?\end{tabular}} & \multicolumn{1}{l}{CM}                                                        & \multicolumn{1}{l}{Training}                                                   & \begin{tabular}[c]{@{}l@{}}Emergency, clinical instructions, \\procedures, and anatomy. For example, \\we learned techniques to safely\\ move them from a bed to a wheelchair \\ or vice versa, protecting from injuries.\end{tabular}                                               & \multicolumn{1}{c}{1}                                                         & \multicolumn{1}{c}{0}       & \multicolumn{1}{c}{0}           & 1                                                            \\ \hline
\end{tabular}
\begin{tablenotes}
\item[1] Elements within the same stanza are conceptually connected. In this context, each question is considered as an individual stanza.
\item[2] To present ENA scores, the unit of analysis will be each caregiver, whereas for the network graph representation, the unit of analysis will be each category.
\item[3] PID refers to Participant ID, indicating each individual caregiver.
\item[4] Category defines the type of question, whether it refers to real-life experiences of caregivers or their training experiences.
\end{tablenotes}
\end{threeparttable}
\end{table*}

The correct formatting and coding of transcribed data are necessary to use ENA to model the connections. In the context of ENA, codes are used to categorize specific aspects or elements of the data being analyzed. Each code represents a distinct concept. We developed our initial coding scheme based on Danzl’s work \cite{danzl2016lot} to analyze the transcription data. Table I presents the schematic version of the codebook, including codes and their corresponding definitions for our study.

To see the coding process, it is helpful to examine an example taken from real coded responses of caregivers during the interview (see Table \ref{tab:Excerpt}).

This sample contains raw data, specifically the caregiver's answers, which are noted in the column titled ``Response".
In this format, each line in the data was associated with an utterance from the transcription data that contained a value corresponding to each code. These values were binary, indicating the presence or absence of each code within the utterances. The code columns (the four columns on the right) show the nodes of the network model: communication, empathy, flexibility, and critical thinking.  Multiple codes could appear together (i.e., code co-occurrence) within the same data line.

In total, we had over 810 utterances that we coded using the four codes seen in Table \ref{tab:coding-scheme}.
Two of the researchers independently coded the records and assigned each utterance to one of these codes. The inter-rater reliability (IRR) via Cohen's kappa \cite{mchugh2012interrater} was $0.863$, and the percentage of agreement was $93.7\%$, showing a strong agreement between the two raters.

\subsection{ENA Method}
In this section, we describe the process by which ENA constructs a network model using the formatted data.
\subsubsection{Constructing adjacency matrices to demonstrate the co-occurrence of codes in each stanza}

In terms of data structuring, the key issue is that there has to be one or more columns that define how to divide the data for analysis. These columns are known as stanzas\cite{shaffer2016tutorial}.
Every line within a stanza is related to one another. In other words, elements that co-occur in a stanza are conceptually linked. In our analysis, we used questions as individual stanzas, and combined all the follow-up questions related to an open-ended question into a single line.  Researchers \cite{chesler2015novel, landauer2013handbook, dorogovtsev2003evolution, lund1996producing, sole2001small} have demonstrated that the frequent co-occurrence of concepts within a specific segment is a good indicator of cognitive connections. Since the ultimate purpose of an ENA is to analyze the structure of connections, finding meaningful connections in the data is critical.

\begin{figure}[H]
    \includegraphics[width=0.5\textwidth]{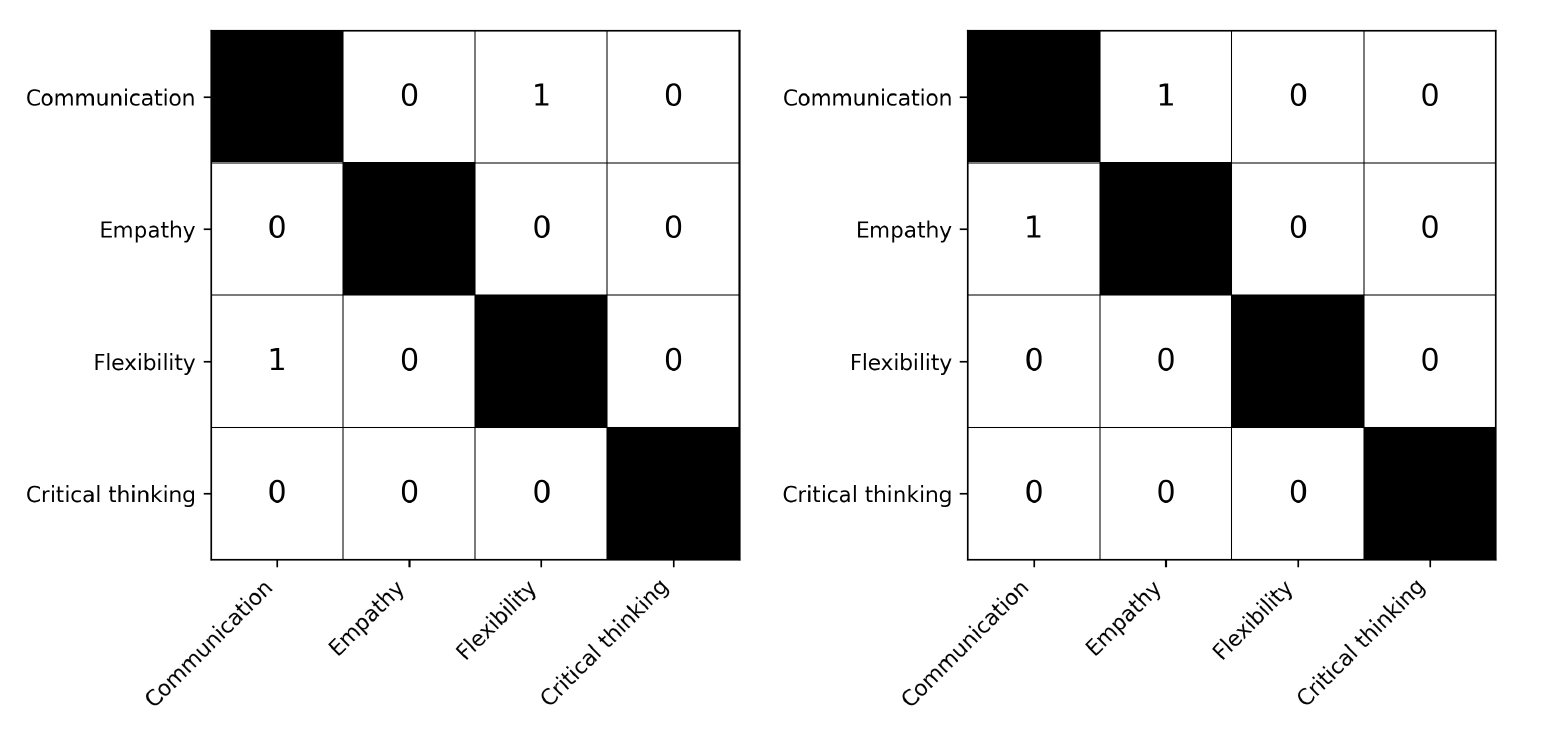} 
    \caption{Adjacency matrices, displaying the co-occurrence of codes, for the first two stanzas presented in Table \ref{tab:Excerpt}. The adjacency matrix has values of zero in diagonal cells, and it is symmetric.}
    \label{fig:Adjacency Matrices}

\end{figure}

In order to achieve these meaningful connections, ENA generates a set of adjacency matrices, each indicating the co-occurrence of codes in a single stanza, to determine the relationships between the objects in the data (See Fig \ref{fig:Adjacency Matrices}). In this matrix, if two codes appear together in the same stanza, the cell corresponding to their intersection is set to one. Conversely, cells representing codes that do not appear in the same stanza are assigned a zero. Since the intersection of codes with themselves does not indicate a relationship between two distinct components, ENA zeroes out the diagonal values. In this way, the first two stanzas in Table \ref{tab:Excerpt}  produce the adjacency matrices shown in Fig \ref{fig:Adjacency Matrices}.

Computing this process individually for each unit, a series of adjacency matrices represent each unit of the dataset.

\subsubsection{Accumulation of adjacency matrices for each unit of analysis}
\begin{figure}[H]
    \centering
    \includegraphics[width=0.23\textwidth]{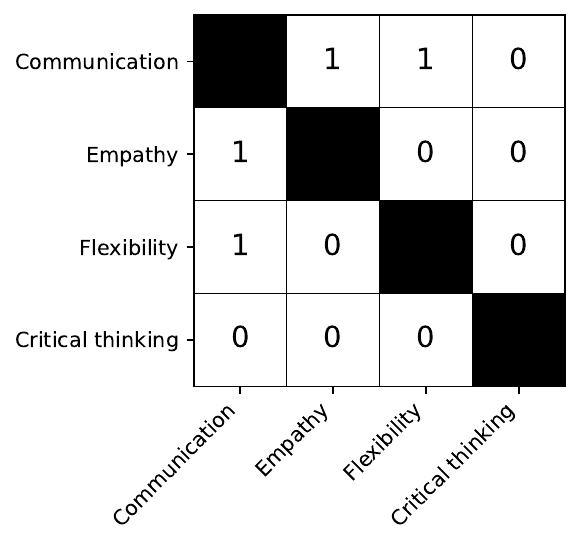} 
    \caption{The cumulative adjacency matrix of two stanzas for a nurse, identified by PID RH, which sums the adjacency matrices presented in Fig \ref{fig:Adjacency Matrices}.}
    \label{fig:Accumulation Matrix}

\end{figure}

To determine the connection structure, ENA aggregates the adjacency matrices of each analysis unit \textit{u} into a single cumulative adjacency matrix, \textit{Cu}. Here, each cell \textit{Cu(i,j)} indicates the count of stanzas where both codes \textit{i} and \textit{j} appeared. In the excerpted dataset shown in Table \ref{tab:Excerpt}, the unit of analysis is each caregiver (Participant ID indicated by PID). Since both stanzas shown in Fig \ref{fig:Adjacency Matrices} belong to the same caregiver, they are combined as illustrated in Fig \ref{fig:Accumulation Matrix} by cell-wise addition.

Each matrix is then transformed into an adjacency vector, \textit{Vu} by copying the cells from the top diagonal of the matrix row by row into a single vector. It's worth noting that because the cumulative adjacency matrix is symmetric, the vector only includes top ( or bottom) diagonal cells. For instance, the vector $[1,1,0,0,0,0]$ would be used to represent the matrix in Fig \ref{fig:Accumulation Matrix}. Thus, these vectors are in a high-dimensional space, \textit{V}, where each dimension denotes a unique pair of two codes. Therefore, the network of connections between objects for each unit is depicted through an adjacency vector in a high-dimensional space, which includes all code co-occurrences accumulated across all stanzas.

\subsubsection{Spherical normalization of adjacency vectors}

In this high-dimensional ENA space, every adjacency vector illustrates the connection pattern of an individual unit, and the length of a vector can be influenced by the total number of stanzas in the analysis unit. To address this issue, ENA spherically normalizes the vectors by dividing each vector \textit{Vu} by its length, ensuring that all vectors have a consistent length. This process normalized the connection strengths to a range of zero to one and accounted for units with different numbers of coded lines. Thus, regardless of the number of stanzas, the normalized vector \textit{Nu} quantifies the relative frequencies of co-occurrence of codes per unit \textit{u}.

\subsubsection{Dimensionality reduction via singular value decomposition}
To interpret and visualize the normalized vectors, ENA applies dimensionality reduction to the high-dimensional space by using singular value decomposition (SVD) \cite{bowman2021mathematical}.
As a result, the original high-dimensional ENA space is rotated, producing orthogonal dimensions that maximize the variance explained by each dimension  \cite{shaffer2016tutorial}. 
For each unit \textit{u} in the dataset, ENA generates a point, \textit{Pu} representing the position of the normalized vector \textit{Nu} within the singular value decomposition framework.

Using the aforementioned approach, we generated two coordinated network representations. The first representation was a set of projected points, referred to as the \textit{ENA score}s, denoting the network position of each unit within the low-dimensional projected space formed by the first two dimensions in the SVD. The second representation was an \textit{ENA network graph} in which nodes corresponded to the codes, and edges depicted the relative occurrence frequency of code pairs for each category, projected into the same low-dimensional space. A deterministic procedure (linear regression) is used to fix node positions on all networks and co-register them with the ENA metrics \cite{bowman2021mathematical}.

As a result, by using a set of fixed node positions, different networks can be directly compared in terms of connected nodes and the strength of their connections.
After employing ENA, we computed the centroid for each network and then applied a two-sample t-test (assuming unequal variance) to evaluate if there were statistical differences between them on both the \textit{X} and \textit{Y} axes. In case of significant differences, we proceeded to examine the specific code connections responsible for these significant differences.
\newline

\section{Results and Discussion}
In this study, we applied Epistemic Network Analysis \cite{bowman2021mathematical, shaffer2016tutorial, shaffer2017epistemic} to our need-finding interview data using the ENA Web Tool (version $1.7.0$) \cite{marquart2018epistemic}.
We created an ENA model to analyze caregivers' behavior in their real-life, daily work settings and compared it with the guidelines provided during their training. The aim was to reveal disparities and identify potential training gaps in order to inform our subsequent design steps for the successful creation of geriatric simulation.

\begin{figure}[H] 
    \centering
    \includegraphics[width=0.5\textwidth]{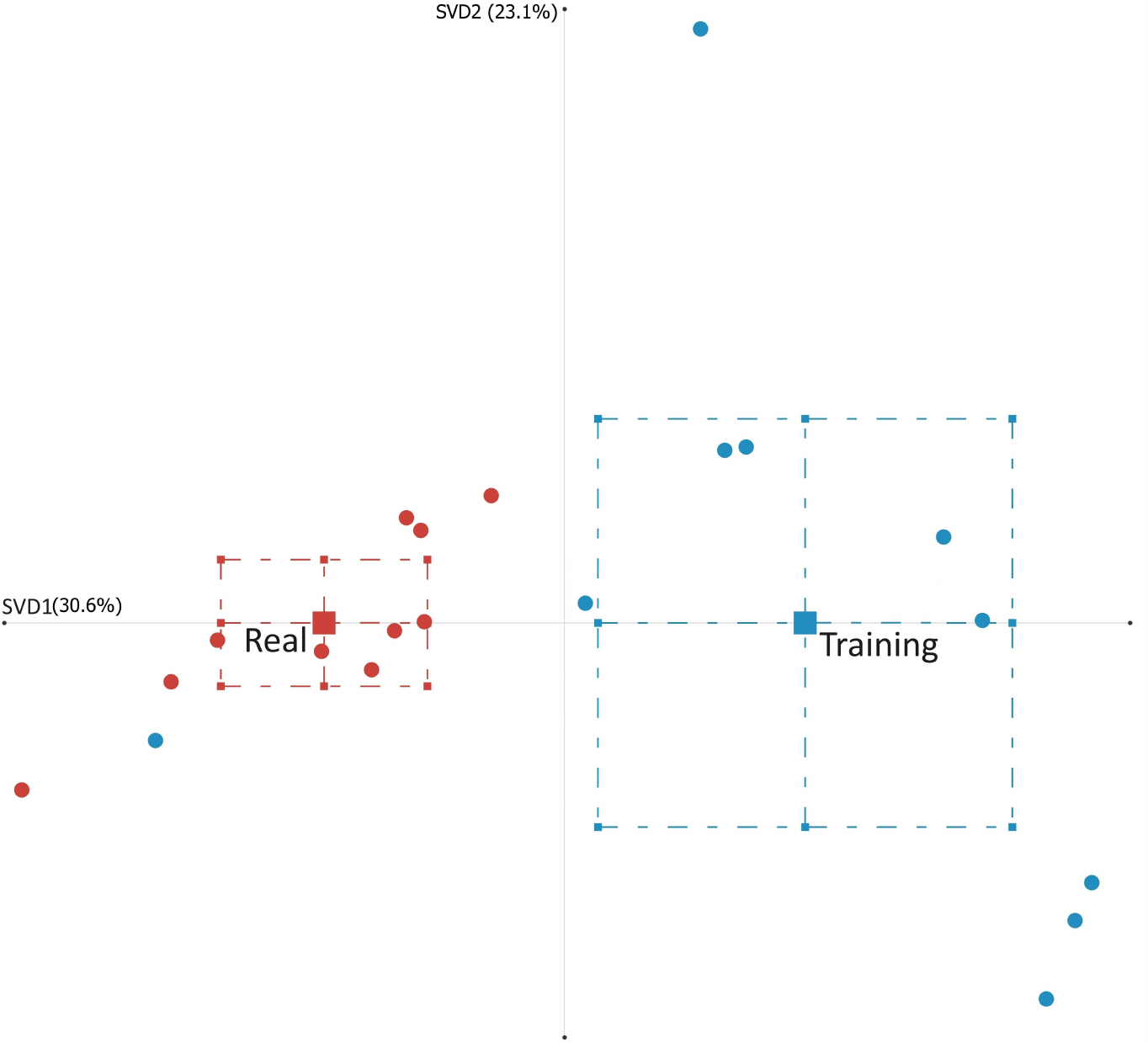} 
    \caption{ENA score plots for Real (red) and Training (blue) response categories. \textit{X-axis} (SVD1) presents a variance of $30.6\%$, \textit{Y-axis} (SVD2) presents a variance of \textit{23.1\%}. Mean and plotted points for each of these two response categories denote a significant difference along the \textit{X-axis}.}
    \label{fig:Scatter Plot}
\end{figure}

Fig \ref{fig:Scatter Plot} shows the plotted points for participants based on their responses in a two-dimensional space generated by SVD, where the first dimension (\textit{X-axis}) represents $30.6\%$ and the second dimension (\textit{Y-axis}) represents $23.1\% $ of the overall data variance.
Red dots represent caregivers' experiences from everyday job activities (Real category), while blue dots correspond to the experiences during training sessions (Training category). Each dot represents the collective responses of individual caregivers (units) to the interview questions. The blue and red squares are mean centroids, and the boxes around these squares are $95\%$ confidence intervals for each dimension. Specifically, the box width indicates the confidence interval along the \textit{X-axis}, whereas the box height depicts the confidence interval along the \textit{Y-axis}.

ENA analysis demonstrated that \textit{30.6\%} of the variance in coding co-occurrences was on the \textit{X} dimension and \textit{23.1\% }of the variance was on the \textit{Y} dimension. We conducted a two-sample t-test (assuming unequal variance) to determine if there was a significant difference between the means of each category. Along the \textit{X-axis }($t(11.06)= -3.05, p=0.01$), the Real response category ($mean=-0.58 \pm 0.60, N=10$) was found to be different from the Training response category ($mean=0.58 \pm 0.81, N=10$), and this difference was statistically significant at the $alpha = 0.05$ level. Cohen’s \textit{d} was equal to $1.63$, also showing a large difference between the two categories. No significant difference was observed along the \textit{Y-axis}.

\begin{figure}[!ht] 

    \centering
   \includegraphics[width=1\linewidth]{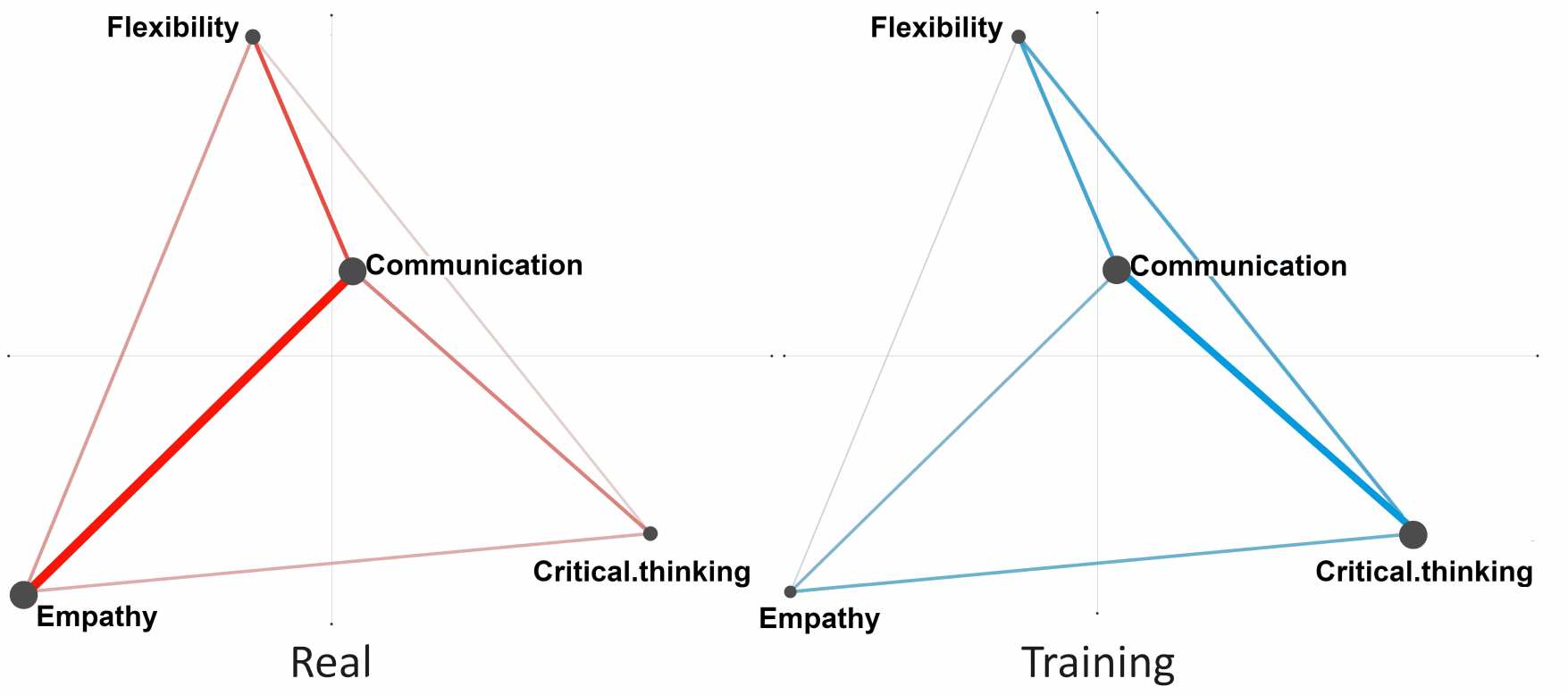}
    \caption{Epistemic network representations of participant responses in an expert need-finding interview process based on Real (red) and Training (blue) categories that show the weighted structure of code connections. The thickness of the edges (lines) indicates the relative strength of co-occurrence between each pair of codes; the size of the nodes (black circles) is proportional to all the connections made by that code.}

    \label{fig:side_by_side}
\end{figure}

We then separately analyzed each network diagram to determine why this statistically different result was found. The codes that moved centroids to the left or right can be identified by examining the strengths of connections. The individual diagrams for Real and Training responses are presented in Fig \ref{fig:side_by_side}.
    
In general, both networks have a relatively strong connection at the communication node. The network for the Real category has relatively light connections in the outer edges between flexibility and empathy, along with critical thinking, resulting in thinner edges. In contrast, a thick, dark red line is apparent between empathy and communication, suggesting a very strong connection. Since the connection between empathy and communication is so strong, it shifts the Real network centroid to the left along the x-axis. The diagram for the Training category shows that the strongest connection is between communication and critical thinking. Since these nodes are located to the right of the x-axis, the strength of this connection shifts the Training network centroid to the right. 

The empathy code was used whenever caregivers showed feelings of understanding of the emotions and thoughts of the patient. The communication code was used whenever caregivers exchanged information and conveyed emotion, even without words \cite{cairns2021empathy}. Therefore, the strong connection between empathy and communication indicates that caregivers in the real workplace express more empathy around certain aspects of communication. In contrast, the size of the empathy node was the smallest in the Training category, and its connections to other nodes were either weak or almost absent. 
The following quotes from two participants also highlight the lack of empathy in training responses and underscore that empathy is a key but missing element of current training practices for nursing assistants:

\begin{itemize}
    \item \textit{``Insufficient preparation has made the lack of empathy the most challenging aspect of working with older patients in this field."}
    \item  \textit{``Empathy helps with the rapport between me and the resident."}
    
\end{itemize} 

Notably, these caregivers talked about their empathic skills and considered the importance of receiving positive feedback, building rapport, and having a good relationship with patients as desired outcomes of their professions. In summary, our presented ENA network models from Real and Training response categories revealed the key elements and their interplay and connections in the geriatric caregiving profession with clear, interpretable illustrations. We learned about the importance of communication in both categories and found areas for improvement in each.

\paragraph{Takeaways and Challenges}
Like any preliminary study, our work had some implications and challenges.
The first challenge was the sample size. While we acknowledge that the important constraint of this exploratory study was a limited sample size, we also recognize that the findings demonstrate the potential of creating ground-breaking participatory design and need-finding considerations with the help of ENA representations for interview data in the community.
Moreover, this work only focused on caregivers' perspectives and did not invite patients or older adults as other key stakeholders of geriatric care to participate in the interviews.  This means caregiver interviews do not necessarily reflect how they communicate with older adults in authentic settings. So, a follow-up study with the inclusion of older adults who receive professional care from nursing assistants will strengthen the study findings. The network comparisons and subtractions to illustrate the different perspectives among older adults and caregivers can be insightful.
In addition, investigating and comparing the findings obtained through QE with those derived from qualitative analysis can be interesting to investigate.
Finally, previous studies have shown that virtual reality simulations can effectively support the clinical readiness of caregivers \cite{shah2022quality, echeverria2019towards}. Therefore, simulated settings have the potential to shift the focus from pure critical thinking to more empathic goals and provide caregivers with authentic experiences of thinking and working in the workplace.
So, the key takeaway is to enhance our future immersive virtual training by including more empathetic modules. Introducing challenging scenarios within the simulation encourages students to carefully consider their decisions and understand their impact on the emotional states of patients. This system could also benefit from feedback mechanisms, providing learners with insights into their performance and the effectiveness of their empathy.

\section{Conclusion}

In this study, we explored the application of network-based modeling techniques for analyzing and visualizing geriatric caregiving attitudes, showcasing the practical implementation of quantitative ethnography principles into interview data. Results from transcriptions of ten participants indicated that caregiver communication was primarily influenced by empathy in real-life situations and by critical thinking during training. The results also highlighted a lack of empathy considerations in caregiver training. These findings equipped us with more insights into the next phase of our work, which involves the design and scenario development of our geriatric care application. We plan to create more realistic training scenarios for caregivers, emphasizing empathy to help caregivers improve the quality of geriatric care. This work has implications in network-based and alternative analytical methodologies in human factors for health. It showcased an example of how epistemic network representations can be leveraged to analyze expert interview data to visually interpret meaningful connections between codes, identify disparities, and assist the research team in making more informed decisions for future design processes.

\section*{Acknowledgment}
 
We would like to thank Dahlia Musa for her assistance with data collection for this study. This work was partially funded by a collaborative grant by the National Science Foundation (2222661, 2222662, and 2222663). The opinions, findings, and conclusions expressed in this study do not necessarily reflect the views of the funding agencies, cooperating institutions, or other individuals.

\balance
\bibliographystyle{IEEEtran}
\bibliography{bibliography}

\begin{thebibliography}{10}
\providecommand{\url}[1]{#1}
\csname url@samestyle\endcsname
\providecommand{\newblock}{\relax}
\providecommand{\bibinfo}[2]{#2}
\providecommand{\BIBentrySTDinterwordspacing}{\spaceskip=0pt\relax}
\providecommand{\BIBentryALTinterwordstretchfactor}{4}
\providecommand{\BIBentryALTinterwordspacing}{\spaceskip=\fontdimen2\font plus
\BIBentryALTinterwordstretchfactor\fontdimen3\font minus \fontdimen4\font\relax}
\providecommand{\BIBforeignlanguage}[2]{{%
\expandafter\ifx\csname l@#1\endcsname\relax
\typeout{** WARNING: IEEEtran.bst: No hyphenation pattern has been}%
\typeout{** loaded for the language `#1'. Using the pattern for}%
\typeout{** the default language instead.}%
\else
\language=\csname l@#1\endcsname
\fi
#2}}
\providecommand{\BIBdecl}{\relax}
\BIBdecl

\bibitem{harrington2020appropriate}
C.~Harrington, M.~E. Dellefield, E.~Halifax, M.~L. Fleming, and D.~Bakerjian, ``Appropriate nurse staffing levels for us nursing homes,'' \emph{Health services insights}, vol.~13, p. 1178632920934785, 2020.

\bibitem{campbell2020interventions}
A.~R. Campbell, D.~Layne, E.~Scott, and H.~Wei, ``Interventions to promote teamwork, delegation and communication among registered nurses and nursing assistants: An integrative review,'' \emph{Journal of Nursing Management}, vol.~28, no.~7, pp. 1465--1472, 2020.

\bibitem{blay2020systematic}
N.~Blay and M.~A. Roche, ``A systematic review of activities undertaken by the unregulated nursing assistant,'' \emph{Journal of Advanced Nursing}, vol.~76, no.~7, pp. 1538--1551, 2020.

\bibitem{wueste2023collaborate}
E.~L.~P. Wueste, ``How to collaborate and not just coexist: an explanatory sequential mixed methods study on the impact of a physician and nurse interprofessional education program on the development of early career pediatrician communication skills and collaborative behaviors once in practice,'' Ph.D. dissertation, University of the Incarnate Word, 2023.

\bibitem{patnaik1999needfinding}
D.~Patnaik and R.~Becker, ``Needfinding: the why and how of uncovering people's needs,'' \emph{Design Management Journal (Former Series)}, vol.~10, no.~2, pp. 37--43, 1999.

\bibitem{lunney2003critical}
M.~Lunney, ``Critical thinking and accuracy of nurses' diagnoses,'' \emph{International Journal of Nursing Terminologies and Classifications}, vol.~14, no.~3, pp. 96--107, 2003.

\bibitem{murray2019new}
M.~Murray, D.~Sundin, and V.~Cope, ``New graduate nurses’ understanding and attitudes about patient safety upon transition to practice,'' \emph{Journal of Clinical Nursing}, vol.~28, no. 13-14, pp. 2543--2552, 2019.

\bibitem{shaffer2016tutorial}
D.~W. Shaffer, W.~Collier, and A.~R. Ruis, ``A tutorial on epistemic network analysis: Analyzing the structure of connections in cognitive, social, and interaction data,'' \emph{Journal of Learning Analytics}, vol.~3, no.~3, pp. 9--45, 2016.

\bibitem{sabo2021self}
K.~Sabo and E.~Chin, ``Self-care needs and practices for the older adult caregiver: An integrative review,'' \emph{Geriatric Nursing}, vol.~42, no.~2, pp. 570--581, 2021.

\bibitem{reinhard2019home}
S.~Reinhard, ``Home alone revisited: Family caregivers providing complex care,'' \emph{Innovation in Aging}, vol.~3, no. Suppl 1, p. S747, 2019.

\bibitem{murfield2020self}
J.~Murfield, W.~Moyle, C.~Jones, and A.~O’Donovan, ``Self-compassion, health outcomes, and family carers of older adults: An integrative review,'' \emph{Clinical Gerontologist}, vol.~43, no.~5, pp. 485--498, 2020.

\bibitem{wolff2020family}
J.~L. Wolff, V.~A. Freedman, J.~F. Mulcahy, and J.~D. Kasper, ``Family caregivers’ experiences with health care workers in the care of older adults with activity limitations,'' \emph{JAMA Network Open}, vol.~3, no.~1, pp. e1\,919\,866--e1\,919\,866, 2020.

\bibitem{faronbi2019caring}
J.~O. Faronbi, G.~O. Faronbi, S.~J. Ayamolowo, and A.~A. Olaogun, ``Caring for the seniors with chronic illness: The lived experience of caregivers of older adults,'' \emph{Archives of Gerontology and Geriatrics}, vol.~82, pp. 8--14, 2019.

\bibitem{singh2015training}
I.~Singh, ``Training and professional development for nurses and healthcare support workers: Supporting foundation for quality and good practice for care of the acutely ill older person,'' \emph{Int Arch Nurs Health Care}, vol.~1, no.~1, p. 007, 2015.

\bibitem{redcross2022}
\BIBentryALTinterwordspacing
R.~Cross, ``Nurse assistant training classes — cna classes,'' 2022, website, Accessed on 02/28/2022. [Online]. Available: \url{https://www.redcross.org/take-a-class/cna/cna-training/cna-classes}
\BIBentrySTDinterwordspacing

\bibitem{emich2019}
\BIBentryALTinterwordspacing
``How nurse staffing affects patient safety and satisfaction,'' October 2019, accessed on 02/28/2022. [Online]. Available: \url{https://online.emich.edu/articles/rnbsn/nurse-staffing-affects-patient-safety-satisfaction.aspx}
\BIBentrySTDinterwordspacing

\bibitem{blazer2016hearing}
D.~G. Blazer, S.~Domnitz, and C.~T. Liverman, ``Hearing health care services: improving access and quality,'' \emph{Hearing Health Care for Adults: Priorities for Improving Access and Affordability}, 2016.

\bibitem{burgdorf2019factors}
J.~Burgdorf, D.~L. Roth, C.~Riffin, and J.~L. Wolff, ``Factors associated with receipt of training among caregivers of older adults,'' \emph{JAMA Internal Medicine}, vol. 179, no.~6, pp. 833--835, 2019.

\bibitem{aksoydan2019training}
E.~Aksoydan, A.~Aytar, A.~Blazeviciene, R.~L. van Bruchem-Visser, A.~Vaskelyte, F.~Mattace-Raso, S.~Acar, A.~Altintas, E.~Akgun-Citak, S.~Attepe-Ozden \emph{et~al.}, ``Is training for informal caregivers and their older persons helpful? a systematic review,'' \emph{Archives of Gerontology and Geriatrics}, vol.~83, pp. 66--74, 2019.

\bibitem{gonzalez2021remotely}
E.~Gonz{\'a}lez-Fraile, J.~Ballesteros, J.-R. Rueda, B.~Santos-Zorroz{\'u}a, I.~Sol{\`a}, and J.~McCleery, ``Remotely delivered information, training and support for informal caregivers of people with dementia,'' \emph{Cochrane Database of Systematic Reviews}, no.~1, 2021.

\bibitem{kamachi2020training}
M.~Kamachi, M.~Owlia, and T.~Dutta, ``Training caregivers to reduce spine flexion using biofeedback,'' in \emph{Advances in Human Factors in Training, Education, and Learning Sciences: Proceedings of the AHFE 2019 International Conference on Human Factors in Training, Education, and Learning Sciences, July 24-28, 2019, Washington DC, USA 10}.\hskip 1em plus 0.5em minus 0.4em\relax Springer, 2020, pp. 241--251.

\bibitem{lindeman2020technology}
D.~A. Lindeman, K.~K. Kim, C.~Gladstone, and E.~C. Apesoa-Varano, ``Technology and caregiving: emerging interventions and directions for research,'' \emph{The Gerontologist}, vol.~60, no. Supplement\_1, pp. S41--S49, 2020.

\bibitem{bramhall2014effective}
E.~Bramhall, ``Effective communication skills in nursing practice,'' \emph{Nursing Standard (2014+)}, vol.~29, no.~14, p.~53, 2014.

\bibitem{wu2021empathy}
Y.~Wu, ``Empathy in nurse-patient interaction: a conversation analysis,'' \emph{BMC Nursing}, vol.~20, no.~1, pp. 1--6, 2021.

\bibitem{acgme2008}
``Iva5d\_educationalprogram\_acgmecompetencies\_ipcs\_explanation.pdf,'' \url{chrome-extension://efaidnbmnnnibpcajpcglclefindmkaj/https://www.acgme.org/globalassets/PDFs/commonguide/IVA5d_EducationalProgram_ACGMECompetencies_IPCS_Explanation.pdf}, 2008, (Accessed on 09/13/2023).

\bibitem{nianih}
``Talking with your older patients | national institute on aging,'' \url{https://www.nia.nih.gov/health/talking-your-older-patients}, 2023, (Accessed on 09/13/2023).

\bibitem{Small2003effectiveness}
\BIBentryALTinterwordspacing
J.~A. Small, G.~Gutman, S.~Makela, and B.~Hillhouse, ``Effectiveness of communication strategies used by caregivers of persons with alzheimer's disease during activities of daily living,'' \emph{Journal of Speech, Language, and Hearing Research}, vol.~46, no.~2, pp. 353--367, 2003. [Online]. Available: \url{https://pubs.asha.org/doi/abs/10.1044/1092-4388\%282003/028\%29}
\BIBentrySTDinterwordspacing

\bibitem{mirzaei2020effectiveness}
A.~Mirzaei~Maghsud, F.~Abazari, S.~Miri, and M.~Sadat~Nematollahi, ``The effectiveness of empathy training on the empathy skills of nurses working in intensive care units,'' \emph{Journal of Research in Nursing}, vol.~25, no.~8, pp. 722--731, 2020.

\bibitem{adler2000relationship}
N.~E. Adler, E.~S. Epel, G.~Castellazzo, and J.~R. Ickovics, ``Relationship of subjective and objective social status with psychological and physiological functioning: Preliminary data in healthy, white women.'' \emph{Health Psychology}, vol.~19, no.~6, p. 586, 2000.

\bibitem{navarro2018engagement}
Y.~Navarro-Abal, M.~J. Lopez-Lopez, and J.~A. Climent-Rodriguez, ``Engagement, resilience and empathy in nursing assistants,'' \emph{Enfermeria Clinica (English Edition)}, vol.~28, no.~2, pp. 103--110, 2018.

\bibitem{mcvicar2003workplace}
A.~McVicar, ``Workplace stress in nursing: a literature review,'' \emph{Journal of Advanced Nursing}, vol.~44, no.~6, pp. 633--642, 2003.

\bibitem{kirca2019relationship}
N.~Kirca and K.~Bademli, ``Relationship between communication skills and care behaviors of nurses,'' \emph{Perspectives in Psychiatric Care}, vol.~55, no.~4, pp. 624--631, 2019.

\bibitem{sagbakken2020adapt}
M.~Sagbakken, R.~Ingebretsen, and R.~S. Spilker, ``How to adapt caring services to migration-driven diversity? a qualitative study exploring challenges and possible adjustments in the care of people living with dementia,'' \emph{PloS One}, vol.~15, no.~12, p. e0243803, 2020.

\bibitem{kang2023toward}
Y.~J. Kang, C.~A. Mueller, J.~E. Gaugler, M.~A. Mathiason~Moore, and K.~A. Monsen, ``Toward ensuring care quality and safety across settings: examining time pressure in a nursing home with observational time motion study metrics based on the omaha system,'' \emph{Journal of the American Medical Informatics Association}, p. ocad113, 2023.

\bibitem{palacio2020resilience}
C.~Palacio~G, A.~Krikorian, M.~J. Gomez-Romero, and J.~T. Limonero, ``Resilience in caregivers: A systematic review,'' \emph{American Journal of Hospice and Palliative Medicine{\textregistered}}, vol.~37, no.~8, pp. 648--658, 2020.

\bibitem{demiris2019problem}
G.~Demiris, D.~P. Oliver, K.~Washington, and K.~Pike, ``A problem-solving intervention for hospice family caregivers: a randomized clinical trial,'' \emph{Journal of the American Geriatrics Society}, vol.~67, no.~7, pp. 1345--1352, 2019.

\bibitem{ploeg2020caregivers}
J.~Ploeg, M.~Northwood, W.~Duggleby, C.~A. McAiney, T.~Chambers, S.~Peacock, K.~Fisher, S.~Ghosh, M.~Markle-Reid, J.~Swindle \emph{et~al.}, ``Caregivers of older adults with dementia and multiple chronic conditions: Exploring their experiences with significant changes,'' \emph{Dementia}, vol.~19, no.~8, pp. 2601--2620, 2020.

\bibitem{shaffer2017quantitative}
D.~W. Shaffer, \emph{Quantitative ethnography}.\hskip 1em plus 0.5em minus 0.4em\relax Lulu. com, 2017.

\bibitem{bowman2021mathematical}
D.~Bowman, Z.~Swiecki, Z.~Cai, Y.~Wang, B.~Eagan, J.~Linderoth, and D.~W. Shaffer, ``The mathematical foundations of epistemic network analysis,'' in \emph{Advances in Quantitative Ethnography: Second International Conference, ICQE 2020, Malibu, CA, USA, February 1-3, 2021, Proceedings 2}.\hskip 1em plus 0.5em minus 0.4em\relax Springer, 2021, pp. 91--105.

\bibitem{shaffer2017epistemic}
D.~Shaffer and A.~Ruis, ``Epistemic network analysis: A worked example of theory-based learning analytics,'' \emph{Handbook of Learning Analytics}, 2017.

\bibitem{shaffer2009epistemic}
D.~W. Shaffer, D.~Hatfield, G.~N. Svarovsky, P.~Nash, A.~Nulty, E.~Bagley, K.~Frank, A.~A. Rupp, and R.~Mislevy, ``Epistemic network analysis: A prototype for 21st-century assessment of learning,'' \emph{International Journal of Learning and Media}, vol.~1, no.~2, pp. 33--53, 2009.

\bibitem{fougt2018epistemic}
S.~S. Fougt, A.~Siebert-Evenstone, B.~Eagan, S.~Tabatabai, and M.~Misfeldt, ``Epistemic network analysis of students' longer written assignments as formative/summative evaluation,'' in \emph{Proceedings of the 8th International Conference on Learning Analytics and Knowledge}, 2018, pp. 126--130.

\bibitem{d2021presentation}
A.~D’Angelo and L.~Ryan, ``The presentation of the networked self: Ethics and epistemology in social network analysis,'' \emph{Social Networks}, vol.~67, pp. 20--28, 2021.

\bibitem{peters2019extending}
E.~E. Peters-Burton, J.~C. Parrish, and B.~K. Mulvey, ``Extending the utility of the views of nature of science assessment through epistemic network analysis,'' \emph{Science \& Education}, vol.~28, no. 9-10, pp. 1027--1053, 2019.

\bibitem{trint2023}
\BIBentryALTinterwordspacing
Trint. (2023) Transcribe video and audio to text: Content editor. Accessed: 2023-08-25. [Online]. Available: \url{https://trint.com/}
\BIBentrySTDinterwordspacing

\bibitem{danzl2016lot}
M.~M. Danzl, A.~Harrison, E.~G. Hunter, J.~Kuperstein, V.~Sylvia, K.~Maddy, and S.~Campbell, ``“a lot of things passed me by”: Rural stroke survivors’ and caregivers’ experience of receiving education from health care providers,'' \emph{The Journal of Rural Health}, vol.~32, no.~1, pp. 13--24, 2016.

\bibitem{mchugh2012interrater}
M.~L. McHugh, ``Interrater reliability: the kappa statistic,'' \emph{Biochemia Medica}, vol.~22, no.~3, pp. 276--282, 2012.

\bibitem{chesler2015novel}
N.~C. Chesler, A.~Ruis, W.~Collier, Z.~Swiecki, G.~Arastoopour, and D.~Williamson~Shaffer, ``A novel paradigm for engineering education: Virtual internships with individualized mentoring and assessment of engineering thinking,'' \emph{Journal of biomechanical engineering}, vol. 137, no.~2, p. 024701, 2015.

\bibitem{landauer2013handbook}
T.~K. Landauer, D.~S. McNamara, S.~Dennis, and W.~Kintsch, \emph{Handbook of latent semantic analysis}.\hskip 1em plus 0.5em minus 0.4em\relax Psychology Press, 2013.

\bibitem{dorogovtsev2003evolution}
S.~N. Dorogovtsev and J.~F. Mendes, \emph{Evolution of networks: From biological nets to the Internet and WWW}.\hskip 1em plus 0.5em minus 0.4em\relax Oxford university press, 2003.

\bibitem{lund1996producing}
K.~Lund and C.~Burgess, ``Producing high-dimensional semantic spaces from lexical co-occurrence,'' \emph{Behavior research methods, instruments, \& computers}, vol.~28, no.~2, pp. 203--208, 1996.

\bibitem{sole2001small}
R.~V. Sol{\'e} \emph{et~al.}, ``The small world of human language.'' \emph{Proceedings. Biological Sciences}, vol. 268, no. 1482, pp. 2261--2265, 2001.

\bibitem{marquart2018epistemic}
C.~Marquart, C.~Hinojosa, Z.~Swiecki, B.~Eagan, and D.~Shaffer, ``Epistemic network analysis (version 1.7. 0)[software],'' \url{https://app.epistemicnetwork.org/}, 2018.

\bibitem{cairns2021empathy}
P.~Cairns, I.~Pinker, A.~Ward, E.~Watson, and A.~Laidlaw, ``Empathy maps in communication skills training,'' \emph{The Clinical Teacher}, vol.~18, no.~2, pp. 142--146, 2021.

\bibitem{shah2022quality}
M.~Shah, A.~Siebert-Evenstone, H.~Moots, and B.~Eagan, ``Quality and safety education for nursing (qsen) in virtual reality simulations: A quantitative ethnographic examination,'' in \emph{Advances in Quantitative Ethnography: Third International Conference, ICQE 2021, Virtual Event, November 6--11, 2021, Proceedings 3}.\hskip 1em plus 0.5em minus 0.4em\relax Springer, 2022, pp. 237--252.

\bibitem{echeverria2019towards}
V.~Echeverria, R.~Martinez-Maldonado, and S.~Buckingham~Shum, ``Towards collaboration translucence: Giving meaning to multimodal group data,'' in \emph{Proceedings of the 2019 Chi Conference on Human Factors in Computing Systems}, 2019, pp. 1--16.

\end{thebibliography}

\end{document}